# A General Framework for Inverse Problem Solving using Self-Supervised Deep Learning
## Validations in Ultrasound and Photoacoustic Image Reconstruction

Jingke Zhang[1], Qiong He[1,2], Congzhi Wang[3,4], Hongen Liao[1], Jianwen Luo[1]*

[1]Department of Biomedical Engineering, School of Medicine, Tsinghua University, Beijing, China
[2]Tsinghua-Peking Joint Center for Life Sciences Department, Tsinghua University, Beijing, China
[3]Paul C. Lauterbur Research Center for Biomedical Imaging, Shenzhen Institutes of Advanced Technology, Chinese Academy of Sciences, Shenzhen 518055, China
[4]National Innovation Center for Advanced Medical Devices, Shenzhen 518055, China
*Email: luo_jianwen@tsinghua.edu.cn

*Abstract*—The image reconstruction process in medical imaging can be treated as solving an inverse problem. The inverse problem is usually solved using time-consuming iterative algorithms with sparsity or other constraints. Recently, deep neural network (DNN)-based methods have been developed to accelerate the inverse-problem-solving process. However, these methods typically adopt supervised learning scheme, which requires ground truths, or labels of the solutions, for training. In many applications, it would be challenging or even impossible to obtain the ground truth, such as the tissue reflectivity function in ultrasound beamforming. In this study, a general framework based on self-supervised learning (SSL) scheme is proposed to train a DNN to solve the inverse problems. In this way, the measurements can be used as both the inputs and the labels during the training of DNN. The proposed SSL method is applied to four typical linear inverse problems for validation, i.e., plane wave ultrasound and photoacoustic image reconstructions, compressed sensing-based synthetic transmit aperture dataset recovery and deconvolution in ultrasound localization microscopy. Results show that, using the proposed framework, the trained DNN can achieve improved reconstruction accuracy with reduced computational time, compared with conventional methods.

## I. Introduction

Solving an inverse problem amounts to discover information about an unknown object of interest from its measurements. Medical image reconstruction is one of its most successful applications, such as computed tomography (CT) [1], magnetic resonance imaging (MRI) [2], positron emission tomography (PET) [3], and ultrasound (US) imaging [4]. These inverse problem-based methods typically start from establishing a measurement model $y = F(x)$ to reflect the relationship between the object $x \in R^m$ and the measurements $y \in R^n$. If the relationship is linear, it can be modeled as:

$$y = Hx + b \quad (1)$$

This work was supported in part by the National Natural Science Foundation of China (61871251, 61801261, 61871263 and 62027901) and Sichuan Science and Technology Program (2019YFSY0048).

where $H \in R^{n \times m}$ is the measurement matrix, $b \in R^n$ denotes the additive noise.

The inverse problem is usually underdetermined (the number of equations $n$ is smaller than the number of unknowns $m$), which leads to lots of possible solutions. Even if in the case where $n = m$, the condition number of $H^{-1}$ is typically very large. Therefore, the solution is very sensitive to noise. To achieve an unique and robust solution for the inverse problem, regularizations should be imposed to introduce the prior knowledge of object in the solving process, such as total variation regularization [5], Tikhonov regularization [6], and wavelet-based sparse regularization [7]. Finally, the inverse problem is solved by using a convex optimization algorithm as follows and shown in Fig. 1(a):

$$\min_{x}\|y - Hx\|_2 + \alpha \|R(x)\|_{1|2} \quad (2)$$

where $\|\cdot\|_2$ denotes the L2 norm, $\|\cdot\|_{1|2}$ denotes the L1 or L2 norm, $R(\cdot)$ is a regularization operator and $\alpha$ is the corresponding weight. Even though the regularized iterative optimization algorithms typically can achieve acceptable results, they are not the best choice to be deployed in real applications, owing to their high computational cost (iterative nature) and difficulty in selection of hyperparameters.

In recent years, deep learning (DL) has become an attractive methodology for medical image analysis for its effectiveness and efficiency [8]. In terms of image reconstruction by solving inverse problems, plenty of DL-based methods have been successfully implemented. The most straight-forward type is to train a deep neural network (DNN) to directly map the measurements (or low-quality reconstruction) to the final high-quality reconstruction non-linearly [9][10], as shown in Fig. 1(b). As an alternative, the deep unrolling methods attempt to combine the conventional model-based methods and data-driven deep learning technique to achieve both accurate and efficient reconstruction [11]–[13]. In these methods, high-quality images obtained using conventional methods [Fig. 1(a)] are needed as training labels, and their quality will determine the performance of trained DNN.

In this work, a self-supervised learning (SSL) approach is used to train the DNN with the guidance of measurement model, as shown in Fig. 1(c). Compared with conventional DNN-based methods, which typically adopt supervised learning scheme, no ground truth of the object is required for training. In many applications, it would be challenging or even impossible to obtain the ground truth, such as the tissue reflectivity function in ultrasound beamforming.

In supervised learning [Fig. 1(b)], the data fidelity term is calculated based on the closeness [such as the L2 norm in (2)] between the output $\hat{x}$ and ground truth $x$ (both in the image domain). Because the ground truth $x$ may be difficult to obtain, the closeness between the $\hat{x}$ and $x$ is equivalent to the closeness between $H\hat{x}$ and measurement $Hx=y$ according to Eq. (1) without regard to the noise. Therefore, in the SSL approach [Fig. 1(c)], the data fidelity term is calculated based on the closeness between the $H\hat{x}$ and measurement $y$ (both in the measurement domain).

In this study, the proposed framework is applied to four typical linear inverse problems of ultrasound and photoacoustic image reconstruction for validation.

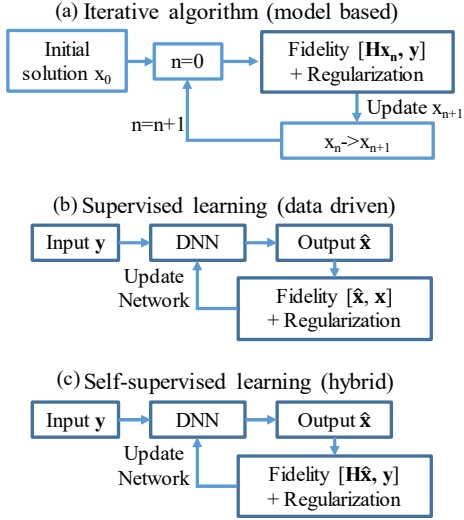

Fig. 1. Diagram of three different inverse problem solving methods.

## II. METHODS

### A. SSL-based Ultrafast Ultrasound Image Reconstruction

The establishment of measurement matrix and the training of the DNN for plane wave ultrasound imaging reconstruction are briefly introduced below, and the details can be found in [14].

A round-trip time-of-flight (TOF) $\tau_{tof}$ can be calculated for the echoes backscattered by each beamforming grid and received by each transducer element. A recording time $\tau_{RF}$ can also be calculated for each sample recorded by transducer elements. The backscattered echo from a beamforming grid would be considered to contribute to the given sample, as long as $\tau_{tof}$ equals $\tau_{RF}$.

A linear measurement matrix $H$ is established according to the TOF relationship described above, to reflect the relationship between the acquired RF channel data $y$ and the desired beamformed data $x$ (without regards to noise),

$$y = Hx \qquad (3)$$

A 12-layers convolutional neural network (CNN) based on an autoencoder architecture is used in this study. Note that the input to the CNN is a low-quality solution $x' = H^T y$ to the inverse problem (obtained using backprojection) rather than the measurement $y$.

As shown in Fig. 1(c), the DNN can be trained by minimizing the mean square root (MSE) loss between the acquired RF channel data $y$ and its recovered version $y'=H\hat{x}$.

Since the problem is ill-posed and underdetermined, regularizations based on prior knowledge of the beamformed signals should be incorporated in the training of DNN. As a typical regularization term used in compressed sensing and sparse regularization methods, the regularization on sparsity of the reconstructed signal in wavelet domain is imposed. Smoothness and sparsity of envelope of the beamformed data in spatial domain are imposed to find a better trade-off between the smoothness (high speckle density) and sparsity (high contrast) of the obtained B-mode image.

When coherent compounding is used in plane wave image reconstruction, two strategies can be used [15]. In this study, the "mapping then sum" strategy was used to obtain the optimal image quality, while another "mapping then sum" strategy could be used to reduce the computational time.

### B. SSL-based recovery of complete STA dataset

The establishment of measurement matrix and the training of the DNN are briefly introduced below, and the details can be found in [16].

According to linear acoustic theory, the echoes received by a specific element of the transducer for an apodized plane wave transmission is a linear combination of received echoes by that element for all STA transmissions with the same apodizations [17]. The linear combination coefficients are the transmit apodization used in PW transmission. Therefore, a linear measurement model can be established as Eq. (3) to reflect the relationship between the received STA data $x \in R^m$ and received PW data $y \in R^n$, where $H \in R^{n \times m}$ is the measurement matrix and corresponds to the applied transmit apodizations.

A 4-layer fully connected network (FCN) is used in this study. Note that the input to the FCN is the slow-time PW data $y$.

As shown in Fig. 1(c), the FCN can be trained by minimizing the mean square root (MSE) loss between the received PW data $y$ and its recovered version $y'=H\hat{x}$.

Since the problem is ill-posed and underdetermined, regularizations based on prior knowledge of the slow-time STA data should be incorporated in the training of FCN. In this study, regularization on sparsity of STA data to be recovered in wavelet domain is imposed.

### C. SSL-based Photoacoustic Image Reconstruction

A linear measurement matrix $H$ is established based on the same TOF relationship as described in Section II.A to reflect the relationship between the acquired RF channel data $y$ and the desired beamformed data $x$. The only difference lies in that there is only one-way TOF $\tau_{tof}$ for each beamforming grid, as the acoustic waves generate simultaneously from the whole pulsed-laser-irradiated region.

Thereafter, a 12-layers convolutional neural network (CNN) based on an autoencoder architecture is used in this study. Note that the input to the CNN is an initial solution $x' = H^T y$ to the inverse problem(obtained using backprojection) rather than measured $y$.

Since the problem is ill-posed and underdetermined, regularizations based on prior knowledge of the beamformed signals should be incorporated in the training of DNN. In this study, the total variation (TV) of the beamformed data is penalized, as it is a typically used regularizer in photoacoustic image reconstruction.

### D. SSL-based Ultrasound localization microscopy

Localizing microbubbles (MBs) is a key step in ultrasound localization microscopy (ULM). In this study, we investigate the feasibility of localizing the MBs (at high concentration) using DNN trained with the proposed framework. The details about the establishment of measurement matrix and the training of the DNN can be found in [18].

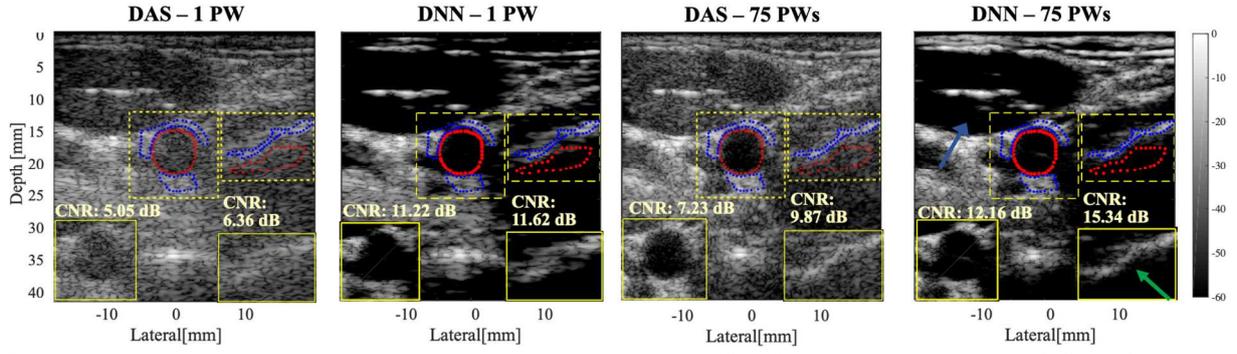

Fig. 2. B-mode images of the common carotid artery in the cross-sectional view of the PICMUS dataset reconstructed using DAS and DNN with 1 and 75 PWs, respectively. The blue arrow indicates that DNN suppresses the background noise effectively. The green arrow indicates that DNN can obtain a clearer tissue structure with higher contrast than DAS.

A linear measurement matrix ***H*** is established to reflect the point spread functions (PSFs) of the system at each imaging grid. The output localization map of the DNN (13-layers CNN) is recovered to the filtered MB image using the forward measurement model, and then used for MSE computation to train the DNN.

*E. Evaluation metrics*

The resolution of reconstructed images is quantified by the full width at half-maximum (FWHM) of the main lobes in the lateral direction and the peak-side-lobe level (PSL), which is defined as the peak value of the first side-lobe. Contrast-to-noise ratio (CNR) is calculated as follows:

$$\text{CNR} = 20 \times \log10\left(\frac{|\mu_{\text{ROI}} - \mu_{\text{BG}}|}{\sqrt{\sigma_{\text{ROI}}^2 + \sigma_{\text{BG}}^2}}\right) \quad (4)$$

where $\mu_{\text{ROI}}$, $\mu_{\text{BG}}$, $\sigma_{\text{ROI}}$, and $\sigma_{\text{BG}}$ are the means and standard deviations of the intensities in the regions of interests (ROIs) and the background regions.

## III. RESULTS

Images reconstructed from the PICMUS dataset [19] with DAS and DNN are shown in Fig. 2. As indicated by the blue arrow, the DNN-based method effectively suppresses the background noise to improve the contrast of tissue structures, compared with DAS using the same number of transmissions. The green arrow indicates that the accuracy of reconstruction improves as the number of transmissions increases. Quantitative results show that the DNN can obtain higher CNR with only 1 PW transmission than DAS using 75 transmissions.

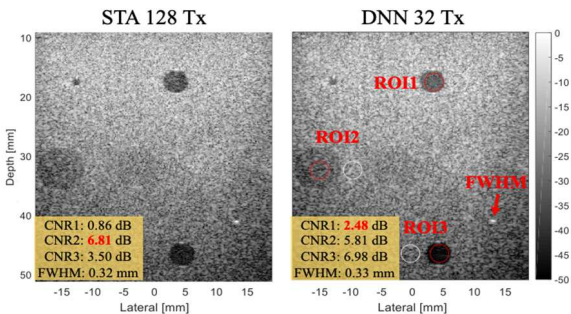

Fig. 3. B-mode STA images reconstructed using different methods. Three ROIs are used to quantitatively evaluate the contrast of these methods. The red arrow indicates the wire target used for FWHM evaluation.

Fig. 3 presents the B-mode images of STA with 128 transmissions and DNN with 32 transmissions. DNN achieves similar lateral resolution to that of STA. In addition, thanks to the high-SNR echoes obtained using apodized PW transmissions, DNN achieves significant improvement on CNR in deep region (ROI 3). Fig. 4 presents the estimated axial and lateral strains from images obtained using STA and DNN. In the axial direction, the DNN achieves similar visual quality to STA in terms of smoothness of strain image, as well as quantitative results in terms of SNR and CNR. In contrast, DNN achieves significantly improved lateral strain, which is in accordance with the SNR and CNR values.

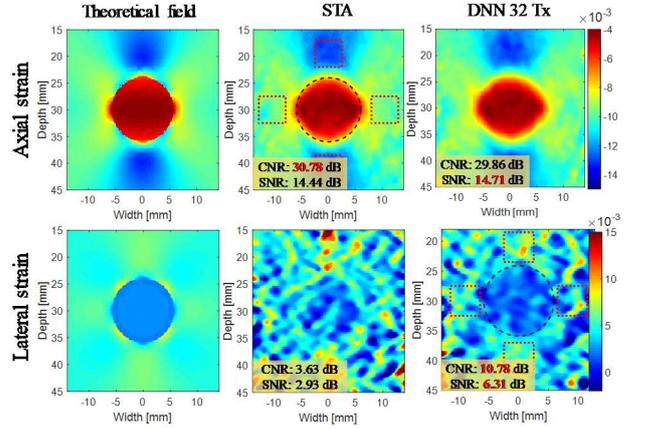

Fig. 4. Axial and lateral strains obtained from different imaging methods, as well as the corresponding theoretical fields. The applied strain is -1%. White Gaussian noise with an SNR of -10 dB is added to STA data and noise with the same power is added to apodized PW data. The black circle and red blocks are ROI and background regions used for CNR calculation, respectively.

Fig. 5 presents reconstructed photoacoustic images using DAS and DNN. The DNN achieves better lateral resolution for a carbon rod placed in the agar phantom, in terms of FWHM and PSL.

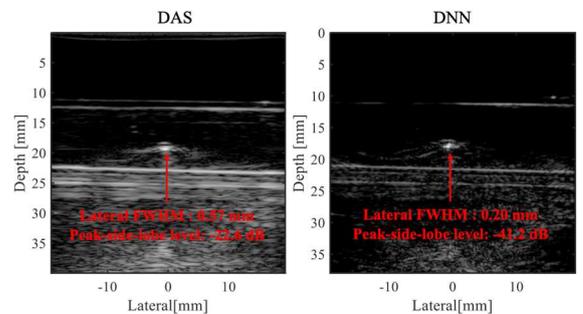

Fig. 5. B-mode images reconstructed using DAS and DNN. Red arrows indicate the carbon rod used for lateral FWHM and PSL quantification.

The temporal resolution is a critical factor in ultrasound imaging. Compared with conventional methods, the reconstruction time for ultrasound and photoacoustic imaging is significantly reduced from 1~5 mins (sparse regularization method) to ~10 ms, and the recovery time of complete STA dataset is significantly reduced from ~1 hour (CS-STA) to ~ 10 seconds.

Fig. 6 presents the ULM images obtained using the conventional method and the proposed method. The trained DNN performs well in the localization of high-concentration MBs, as indicated by the white arrows. Some microvessels, which are not detected by the conventional method, can be clearly visualized by the proposed method.

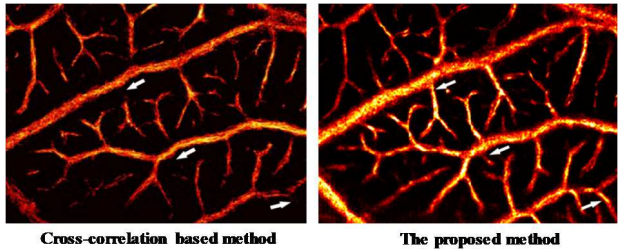

Fig. 6. The ULM images obtained using conventional cross-correlation – based method and a DNN trained using the proposed framework.

## IV. DISCUSSION

In this study, the feasibility of the proposed framework for inverse problem solving was validated in ultrasound and photoacoustic image reconstruction. Results demonstrate that DNN trained using self-supervised learning can achieve improved reconstruction accuracy with reduced computational time, compared with conventional methods.

The proposed framework can be extended to solve other inverse problems or image reconstruction problems, such image reconstruction of CT, MRI and PET, as well as deconvolution in fluorescence localization microscopy [20] and ultrasound localization microscopy [21].

The measurement matrix is a bridge to connect the measurement data and the signals to be reconstructed, and it also plays an important role in loss computation during the training of DNN. Therefore, it is vital to ensure the accuracy of established measurement matrix.

In ultrasound and photoacoustic image reconstruction, the input of DNN is not the measurement [with size of 1300 × 128, (axial × lateral)], but is the low-quality solution to inverse problem obtained using backprojection method. In contrast, for the recovery of STA data, the input of DNN is exactly the measurement (a vector with length of 32).

The choice of input data depends on the complexity of measurement model. According to the measurement model $y = Hx$, each element of $y$ is a linear combination of $x$. Sometimes, the relationship is too complex for DNN to learn. In this case, it is recommended to provide low-quality initial solution to allow a local feature extraction. As the inverse problem is typically ill-posed and underdetermined, regularizations based on the prior knowledge of the signals to be reconstructed should be imposed as loss function during training to improve the accuracy of reconstruction.

## V. CONCLUSION

In this study, a general framework based on self-supervised learning is proposed to train a deep neural network to solve the inverse problems. The feasibility of the proposed framework was validated in four typical linear inverse problems. Results show that, by using the proposed framework, the trained DNN can achieve improved reconstruction accuracy with reduced computational time, compared with conventional methods. In conclusion, the proposed method can achieve a better trade-off between the image quality and reconstruction time, which may be helpful for real-time applications of medical imaging reconstruction.